\documentclass[prl,twocolumn,showpacs]{revtex4}
\usepackage{graphicx}

\newcommand{\paper}{Letter}
\newcommand{\species}[1]{\emph{#1}}
\newcommand{\Ecoli}{\species{E.\ coli}}

\newcommand{\latin}[1]{\emph{#1}}

\newcommand{\ie}{\latin{i.}$\,$\latin{e.}}

\newcommand{\growth}{\mathrm{gr}}
\newcommand{\vgr}{v_{\growth}}
\newcommand{\ci}{c_i}
\newcommand{\va}{v_\alpha}
\newcommand{\mymin}{\mathrm{min}}
\newcommand{\vamin}{\va^\mymin}
\newcommand{\vglc}{v_{\mathrm{glc}}}
\newcommand{\myb}{B}
\newcommand{\myba}{\myb_\alpha}
\newcommand{\mybbio}{\myb_\alpha}
\newcommand{\mybdrain}{\myb^*}

\begin{document}

\title{Duality, thermodynamics, and the linear programming problem\\
in constraint-based models of metabolism}

\author{Patrick B. Warren}
\author{Janette L. Jones}
\affiliation{Unilever R\&D Port Sunlight, Bebington, Wirral, CH63 3JW, UK.}

\date{June 11, 2007 --- 2nd revision}

\begin{abstract}
It is shown that the dual to the linear programming problem that
arises in constraint-based models of metabolism can be given a
thermodynamic interpretation in which the shadow prices are
chemical potential analogues, and the objective is to minimise free
energy consumption given a free energy drain corresponding to growth.
The interpretation is distinct from conventional non-equilibrium
thermodynamics, although it does satisfy a minimum entropy production
principle.  It can be used to motivate extensions of constraint-based
modelling, for example to microbial ecosystems.
\end{abstract}

\pacs{05.70.Ln, 82.60.-s, 87.16.-b}


\maketitle

In biology, the metabolism of an organism provides energy and raw
materials for maintenance and growth.  As such, an interesting and
important question concerns the application of thermodynamics to
metabolic reaction networks \cite{XBeard, XHJBH, XvSMLMP, KPH}.  For
example, Prigogine and Wiame suggested a long time ago that an
organism's metabolism might be governed by a minimum entropy
production (MEP) principle \cite{XMEP}.  From the physical point of
view, a metabolic reaction network is an excellent example of a system
in a non-equilibrium steady state, since one can usually assume that
the metabolite concentrations are unchanging after a short transient
relaxation period.  The appropriate generalisation of thermodynamics
and statistical mechanics to non-equilibrium steady-states is a large
field \cite{GMbook}, which continues to attract attention to this
present day \cite{Xtra}.  In this \paper, we show that a novel
thermodynamic interpretation can be given to the dual linear
programming problem which arises in constraint-based models of
metabolism.  The resulting interpretation is rigorously defined, and
uniquely determined by the mathematics.  It is closely analogous to,
but distinctly different from, conventional non-equilibrium
thermodynamics.  We also show that it satisfies an MEP principle
similar to that proposed by Prigogine and Wiame.

Constraint-based modelling (CBM) of metabolic networks has been
pioneered by Palsson and co-workers \cite{cbmbook}.  In a typical
application, described in more detail below, the steady-state
assumption is combined with a target function to make a linear
optimisation or linear programming (LP) problem.  The LP variables are
the fluxes through the various reactions that comprise the network,
and the LP constraints arise from basic considerations of
stoichiometry and from the reversibility or otherwise of the
reactions.  The LP objective function is biologically motivated, for
example a `growth' reaction is commonly inserted, and the target is to
maximise flux through this reaction to correspond to maximal growth
rate.  CBM has been applied to microorganisms from all three domains
of life \cite{XiJR904, XiND750, XiAF692}, and has been remarkably
successful in predicting phenotypic behaviour \cite{IEP, XZhang, TS}.

\begin{figure}
\begin{center}
\includegraphics{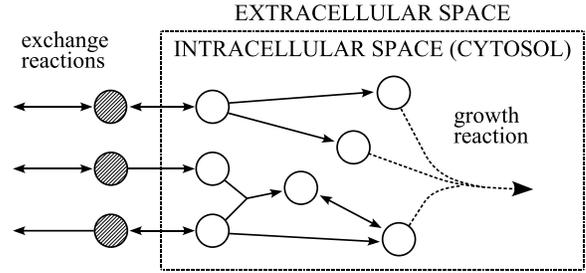}
\end{center}
\caption[?]{Schematic metabolic network for a prokaryote like \Ecoli,
showing intracellular metabolites (open circles), extracellular
metabolites (hatched circles), internal and exchange reactions
(arrows), and a growth reaction (dashed arrow).\label{fig1}}
\end{figure}

\begin{table*}
\begin{ruledtabular}
\begin{tabular}{llll}
 & Primal & Dual & CS inequalities \\
\hline\\[-9pt]
variables & fluxes, $\va$ & chemical potentials, $\mu_i$ \\[3pt]
flux balance / thermodynamics & 
$\sum_\alpha S_{i\alpha}\va=0$ & 
$\myba = \sum_i \mu_i S_{i\alpha}$ \\[3pt]
reversible reactions & $\va$ unlimited & $\myba=0$ \\[3pt]
irreversible reactions & $\va\ge0$ & $\myba\le0$ &
$\va\myba\le0$ \\[3pt]
growth reaction & $\vgr\ge0$ & $\mybbio\le-\mybdrain<0$ &
$\vgr(\mybbio+\mybdrain)\le0$ \\[3pt]
limited exchange reactions & $\va\ge-\vamin$ & $\myba\le0$ &
$(\va+\vamin)\myba\le0$ \\[3pt]
objective function & $z=\vgr$ & 
$w=\sum_\alpha\vamin|\myba|/\mybdrain$\\[3pt]
\end{tabular}
\end{ruledtabular}
\caption[?]{The primal and dual linear programming problems in
constraint-based models of metabolism.  In the dual objective
function, the sum is over the limited exchange reactions only.  The
complementary slackness (CS) inequalities are saturated (\ie\ ${}=0$)
at optimality, where also $\max\,z=\min\,w$. The shadow prices for the
primal problem are given by $-\mu_i/\mybdrain$.\label{table1}}
\end{table*}

Mathematically, every LP problem has a unique dual \cite{lpbook}.  It was in
determining the dual to the CBM LP problem that we noticed a striking
analogy to non-equilibrium thermodynamics.  Let us start therefore
with a general discussion of LP duality, before specialising to the
case of CBM.  We recall that the basic or primal LP problem, in
standard form, is to maximise an objective function
$z=\sum_{\alpha=1}^{n} a_\alpha x_\alpha$ given $\sum_{\alpha=1}^{n}
A_{i\alpha} x_\alpha = b_i$ ($i = 1 \dots m$, $m < n$), where the
$x_\alpha\ge0$ are variables, the $a_\alpha$ are coefficients,
$A_{i\alpha}$ is a matrix, and the $b_i$ are constants (we use Greek
and Roman indices to emphasise that different components live in
different vector spaces).  The dual problem is then to minimise an
objective function $w = \sum_{i=1}^{m} \pi_i b_i$ subject to
$\sum_{i=1}^{m} \pi_i A_{i\alpha} \ge a_\alpha$, with no restriction
on the sign of dual variables $\pi_i$.  The LP strong duality theorem
guarantees that $\max\,z = \min\,w$, provided both problems have
optimal solutions.  In addition, at optimality, `complementary
slackness' (CS) conditions hold.  To formulate these, first define the
`slack' in the inequalities in the dual problem to be $y_\alpha =
\sum_{i=1}^{m} \pi_i A_{i\alpha} - a_\alpha$.  The CS conditions
state that the inequalities $x_\alpha y_\alpha \ge 0$ are saturated
(\ie\ ${} = 0$) at optimality, and only at optimality.

In many applications of LP, the dual problem can be given an economic
interpretation, which has led to the dual variables being generically
known as `shadow prices'.  We note that shadow prices can be obtained
directly from the solution to the primal problem \cite{lpbook}, so the
dual problem need never be explicitly formulated.  This may be the
reason why the remarkably simple structure of the dual problem in CBM
has not been described before.  The use of shadow prices in CBM was
pioneered by Varma and Palsson to assess efficiencies in a model of
the central metabolism of \Ecoli\ \cite{VP1and2}.

Now let us turn to the LP problem in CBM.  We start with the set of
chemical rate equations that describe the metabolic reaction network,
$d\ci/dt = \sum_\alpha S_{i\alpha} \va$, where the $\ci$ are
metabolite concentrations, the $\va$ are reaction velocities or
fluxes, and $S_{i\alpha}$ is a stoichiometry matrix giving the number
of moles of the $i$th metabolite involved in the $\alpha$th reaction.
Making the steady-state assumption, the chemical rate equations reduce
to a set of flux-balance conditions $\sum_\alpha S_{i\alpha} \va=0$.
At this point, in CBM, attention shifts from the metabolite
concentrations to the reaction fluxes.  From this point of view, the
flux-balance conditions become a set of linear constraints on the
$\va$.  In addition, one usually imposes the `thermodynamic'
constraint that $\va\ge0$ if a reaction is irreversible.

In modern approaches \cite{cbmbook}, the reactions in the network are
elementally and charge-balanced.  To make the network `do' something,
two kinds of imbalanced reactions are typically added.  The first, as
mentioned already, is a growth reaction.  This reaction drains the
endpoints of metabolism in the appropriate ratios and represents the
combined effect of the biochemistry subsequent to metabolism.  The
flux through the growth reaction (the growth rate) will be labelled
$\vgr$.  The second type of imbalanced reaction is an `exchange'
reaction, which represents the exchange of an extracellular metabolite
with the environment (the model additionally includes transporter
reactions which allow extracellular metabolites to enter and leave the
intracellular environment).  The exchange reactions enable the uptake
of food substrates, trace minerals, dissolved gases, and vitamins; and
the discharge of metabolic waste products.  By convention, a positive
(negative) flux through an exchange reaction represents the discharge
(uptake) of the corresponding metabolite.  Exchange reactions may be
reversible, or irreversible if discharge only is possible.  A special
case arises when one wishes to represent \emph{limited} availability,
for example of a food substrate.  In this case the exchange flux is
allowed to become negative to a limited extent, thus $\va\ge-\vamin$
with $\vamin>0$ representing a `cap' on the (negative) reaction flux.
The value of $\vamin$ is typically empirically determined to agree
with experimentally measured uptake rates.  Fig.~\ref{fig1} shows
schematically how the exchange reactions and the growth reaction are
connected into the rest of the metabolic network.  For high accuracy
work, an $\mathrm{ATP}\to\mathrm{ADP}$ maintenance reaction with a
specified flux is sometimes included in the model \cite{XiAF692}.  We
omit this here although it can easily be accommodated with a small
extension to the formalism.

The LP problem in CBM is then to find values for the fluxes $\va$
which maximise $\vgr$ subject to the above constraints.  This is
summarised in Table \ref{table1}.  Usually it is the limited
availability of substrates through the exchange reactions that
prevents the problem being unbounded.  Technically the LP problem is
not quite in the standard form but it does not take much to make it
so.

We formulate the dual to the above (primal) LP problem following the
textbook approach described above.  After some straightforward
simplifications, the following picture emerges.  Each metabolite has
an associated shadow price $\pi_i$ which is unrestricted in sign.
Each reaction has an associated constraint, of the form
$\sum_i\pi_iS_{i\alpha}=0$ for reversible and unlimited exchange
reactions, $\sum_i\pi_iS_{i\alpha}\ge0$ for irreversible and limited
exchange reactions, and $\sum_i\pi_iS_{i\alpha}\ge1$ for the growth
reaction.  The objective function is $w = \sum_{i\alpha}\pi_i
S_{i\alpha}\vamin$ where the sum is over the limited exchange
reactions only.  The LP problem is to find values for the shadow
prices $\pi_i$ which \emph{minimise} this objective function subject
to the constraints.  Note that, although the $\vamin$ appear in the
dual objective function $w$, these are numerical constants common to
both the primal and dual problems.  The actual fluxes $v_\alpha$ do
not feature in the dual problem.

\begin{figure}
\begin{center}
\includegraphics{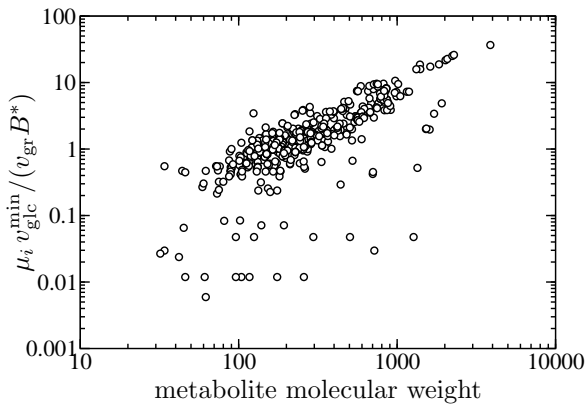}
\end{center}
\caption[?]{Chemical potential analogues (shadow prices) plotted as a
function of metabolite molecular weight, for the intracellular
metabolites with $\mu_i>0$ in a genome-scale constraint-based model of
\Ecoli\ \cite{note-model}.  A normalising factor is included to make
the $\mu_i$ dimensionless.\label{fig2}}
\end{figure}

We now show that the dual problem admits a striking thermodynamic
interpretation.  The motivation is the standard expression for the
free energy change in a chemical reaction, or reaction affinity,
$A_\alpha = \sum_i\mu_iS_{i\alpha}$, where the $\mu_i$ are chemical
potentials \cite{GMbook}.  The similarity between this, and the rules
for formulating the dual LP problem above, make it natural to
interpret the shadow prices as chemical potential analogues.  To aid
the interpretation, we rescale the dual problem by a factor
$-\mybdrain<0$, set $\mu_i=-\pi_i\mybdrain$, and write
$\myba=\sum_i\mu_iS_{i\alpha}$ as the analogue of reaction affinity.
The resulting thermodynamic formulation of this rescaled dual LP
problem is summarised in Table~\ref{table1}.  We have introduced
$\myba$ to distinguish our interpretation from conventional
non-equilibrium thermodynamics; in general $\myba\ne A_\alpha$, as
explained in more detail below.

Let us discuss the thermodynamic interpretation in a bit more depth.
We see that the constraints assert that $\myba=0$ for a reversible
reaction, and $\myba\le0$ for an irreversible reaction.  These are
precisely in accord with equilibrium chemical thermodynamics.  In
addition we interpret the fact that $\myba\le-\mybdrain<0$ for the
growth reaction to mean that a minimum free energy drain equal to
$\mybdrain$ is required for growth.  The magnitude of $\mybdrain$ sets
the overall energy scale, and can be arbitrarily chosen.  Finally, in
the rescaled dual LP problem the objective is to minimise
$w\mybdrain=\sum_\alpha \vamin |\myba|$, in other words a weighted sum
of the free energy consumption associated with the limited exchange
reactions.  At optimality, one has $\max\,z=\min\,w$, hence the growth
rate is easily calculated from the solution to the thermodynamic LP
problem using $\vgr = {\sum_\alpha\vamin|\myba|} / {\mybdrain}$.

We now show that the formalism satisfies an MEP principle.  To derive
this, we consider the internal entropy production due to the chemical
transformations, $T\dot S=-\sum_\alpha\va\myba$.  The sum excludes the
exchange reactions since the flux balance condition implies the total
entropy production $\sum_\alpha\va\myba\equiv0$ when the sum is over
\emph{all} reactions.  It is straightforward to show that $z\le
T\dot S/\mybdrain\le w$.  Thus, at optimality, the entropy production
is `pinched' between the two objective functions.  Alternatively, for
a fixed growth rate $\vgr$, one has $\vgr\mybdrain\le T\dot S$.  Since
the minimum value is attained at the combined solution of the primal
and dual problems, this gives the desired MEP principle.

What does the dual solution look like for a constraint-based model of
metabolism?  To give an example, we computed the chemical potential
analogues for a genome-scale model of the metabolism of \Ecoli\
growing aerobically on a glucose `minimal medium', for which uptake of
extracellular glucose is the limiting exchange reaction
\cite{note-model}.  Lack of space precludes a detailed discussion of
the results, but we find that the vast majority of chemical potentials
are positive and there is a broad distribution over several decades of
magnitude.  An interesting observation is that the chemical potentials
increase with increasing molecular complexity.  This is shown in
Fig.~\ref{fig2}, using molecular weight (discounting metal ions) as a
stand-in for molecular complexity.  This correlation arises because
the chemical potential of a complex molecule is given approximately by
the sum of the chemical potentials of its constituent parts.  This in
turn follows from the CS conditions which imply $\myba =
\sum_i\mu_iS_{i\alpha} = 0$ for all reactions with a flux $\va\ne0$
(see further discussion below).

Let us discuss our findings in a wider context.  Our results show that
the dual to the CBM LP problem has a thermodynamic interpretation in
which the dual variables are analogous to chemical potentials.  The
rules to formulate this thermodynamic LP problem are summarised in
Table~\ref{table1}.  In principle we could strip away the CBM
`scaffolding' and let the thermodynamic LP problem stand on its own,
since LP duality guarantees this is equivalent to solving the original
(primal) LP problem.  Such a viewpoint motivates a number of
interesting questions.

Firstly, a technical point arises since the primal LP problem is often
\emph{degenerate}, in the sense that alternative optimal flux
distributions exist \cite{MS}.  This reflects the fact that multiple
pathways may exist in the metabolism.  But this is not a serious
problem, for the dual problem will be similarly degenerate but the
strong duality theorem and the CS conditions still hold, allowing one
to move from a solution of the dual problem to a solution of the
primal problem, and \latin{vice versa}.

A more serious discussion point concerns the relationship to
conventional non-equilibrium thermodynamics.  For the reversible
reactions, $\myba=0$ is a constraint.  For the irreversible reactions,
the CS conditions (Table \ref{table1}) show that $\myba=0$ if there is
a flux ($\va>0$) through a reaction, and $\myba<0$ only if there is no
flux ($\va=0$) through a reaction (except for the growth reaction
where we expect $\vgr>0$ and hence $\myba=-\mybdrain$).  This presents
a sharp contrast to conventional non-equilibrium thermodynamics where
a flux through a reaction ($\va>0$) is associated with a (negative)
affinity driving force $A_\alpha<0$.  This clearly demonstrates that
$\myba\ne A_\alpha$, and the thermodynamics described in Table
\ref{table1} is not simply the same as conventional non-equilibrium
thermodynamics.  We must therefore regard the rules described in
Table~\ref{table1} as describing a \emph{novel} but \emph{tightly
constrained} thermodynamics for the CBM class of problems, derived
from the (unique) dual to the primal LP problem.  Whether the close
analogy to \emph{equilibrium} chemical thermodynamics (and the
unexpected appearance of an MEP principle) is indicative of deeper
principles or not remains a problem for future investigation.  It
would, for example, be an interesting exercise to compare the \Ecoli\
shadow prices with what is known about the thermodynamic metabolic
state of this organism \cite{XHJBH}. We should emphasise that our MEP
principle is couched in terms of $\myba$ and not $A_\alpha$, and
therefore our results do not constitute a proof of the original
proposition of Prigogine and Wiame \cite{XMEP}.

Another interesting remark is that the choice in the primal LP problem
to maximise the flux through a growth reaction seems to be `pure
biology'.  Experiments demonstrate that this works well under
controlled conditions \cite{IEP, XZhang}, and it can be supported by
examining population dynamics for continuous culture growth in a
chemostat \cite{PirtBook}.  Other choices could and perhaps should be
made in different circumstances \cite{Segre, VGP, WP}.  In terms of
the thermodynamic LP problem, this biologically-motivated component is
translated into the existence of a growth reaction with a minimal free
energy drain $\mybdrain$.  We could turn this observation to our
advantage, to suggest extensions to the CBM approach which are
perhaps unobvious in the primal LP problem.  Consider for example
metabolism in a microbial \emph{ecosystem}, comprising multiple
species which share a pool of common extracellular metabolites.  The
obvious generalisation of the thermodynamic LP problem is to include a
growth reaction with a minimal free energy drain $\mybdrain$ for each
organism, and seek to minimise the free energy consumption \emph{of
the ecosystem} through the exchange reactions of the extracellular
metabolites.  Of course LP duality means there is a corresponding
primal model (in this case the primal objective function becomes a
weighted sum of growth rates \cite{XSDHPLLS}).  Further exploration of
this we leave to future work.

We thank J.\ D.\ Trawick and S.\ J.\ Wiback of Genomatica, Inc., for
useful correspondence; M.\ E.\ Cates, W.\ C.\ K.\ Poon, and P.\ R.\ ten
Wolde for a critical reading of the manuscript; and B.\ \O.\ Palsson
and his group for helpful discussions and generously providing the
\Ecoli\ model.  More details of our derivations, and more examples,
will be presented in a longer publication currently in preparation.


\begin{thebibliography}{23}
\bibitem[{not()}]{note-model}
\bibinfo{note}{The model is based on `iJR904' described in Ref.~\cite{XiJR904}.
  We set $\vglc^\mymin = 10 \,
  \mathrm{mmol}/\text{gram-dry-weight}/\mathrm{hr}$ for which we find a growth
  rate $\vgr = 0.957/\mathrm{hr}$.}
\end{thebibliography}

\end{document}